\documentclass[a4paper]{article}

\usepackage{spconf}
\usepackage{amsmath,graphicx}
\usepackage{amssymb,amsmath,bm}
\usepackage{textcomp}
\usepackage{algorithm}
\usepackage{algpseudocode}
\usepackage{float}
\usepackage{multirow}
\usepackage[justification=centering]{caption}
\usepackage[caption = false]{subfig}

\usepackage{tikz,placeins}
\tikzstyle{every picture}+=[font=\rmfamily\it\bfseries\large]

\newcommand{\specialcell}[2][c]{%
     \begin{tabular}[#1]{@{}c@{}}#2\end{tabular}}

\title{Power-law nonlinearity with maximally uniform distribution criterion for
improved neural network training in automatic speech recognition}
\name{{Chanwoo Kim, Mehul Kumar, Kwangyoun Kim, and Dhananjaya Gowda} 
  \thanks{Thanks to Samsung
Electronics for funding this research. The authors are thankful to 
Executive Vice President Seunghwan Cho and speech processing Lab.
members at Samsung Research.}}

\address{Samsung Research \\
  {\small \tt \{chanw.com, mehul3.kumar, ky85.kim, d.gowda\}@samsung.com }}
\begin{document}
%
\maketitle
\begin{abstract}
In this paper, we describe the Maximum Uniformity of Distribution 
(MUD) algorithm with the power-law nonlinearity. In this approach, 
  we hypothesize that neural network training will become more stable
 if feature distribution is not too much skewed.
We propose two different types of MUD approaches: power function-based MUD 
and histogram-based MUD. In these approaches, we first obtain the mel filterbank
coefficients and apply nonlinearity functions for each filterbank channel.
With the power function-based MUD, we apply a power-function based
nonlinearity where power function coefficients are chosen to maximize the 
likelihood assuming that nonlinearity outputs follow the uniform distribution.
With the histogram-based MUD, the empirical Cumulative Density Function (CDF) 
from the training database is employed to transform the original distribution
 into a uniform distribution.
In MUD processing, we do not use any prior knowledge (e.g. logarithmic
relation) about the energy of the incoming signal and the perceived intensity 
by a human. Experimental results using an end-to-end speech recognition system
demonstrate that power-function based MUD shows better result than the 
 conventional Mel Filterbank Cepstral Coefficients (MFCCs).  On the LibriSpeech
  database, we could achieve 4.02 \% WER on {\tt test-clean} and 13.34 \% WER on
  {\tt test-other } without using any Language Models (LMs). The major
  contribution of this work is that we developed a new algorithm for designing
  the compressive nonlinearity in a {\it data-driven} way, which is much more
  flexible than the previous approaches and may be extended to other domains as
  well.
\end{abstract}
%

%

%
%
\section{Introduction}
After the breakthrough of deep learning technology
\cite{Seltzer2013DNNAurora4, Yu2013FeatureLearningDNN,
V_Vanhoucke_Deep_Learning_NIPS_Workshop_2011,
G_Hinton_IEEE_Signal_Process_Mag_2012, h_hadian_interspeech_2018_00, 
s_karita_interspeech_2019_00, c_chiu_icassp_2018_00, r_prabhavalkar_interspeech_2017_00}, 
speech recognition accuracy has
improved dramatically. Recently, speech recognition systems
are widely used not only in smart phones and Personal Computers
(PCs) but also in standalone devices in far-field environments.
Examples include voice assistant systems such as Amazon Alexa
, Google Home \cite{c_kim_interspeech_2017_00, B_Li_INTERSPEECH_2017_1}, and
Samsung Bixby \cite{samsung_bixby}.

In the era of deep neural networks, it has been frequently
observed that the amount and coverage of the training data 
seem to be one of the most important factors to obtain better 
speech recognition accuracy 
\cite{h_soltau_interspeech_2017_00, a_narayanan_slt_2018_00}.
However, it is very difficult to gather sufficient amount of
transcribed data from various domains. To overcome this problem,
data augmentation has been very popular these days 
\cite{w_hartmann_interspeech_2016_00, x_cui_taslp_2015_00, s_park_interspeech_2019_00, c_kim_icassp_2018_00, 
c_kim_interspeech_2019_00}. Small Power Boosting (SPB) technique
may be considered as a variation of data augmentation techniques
\cite{C_Kim_ASRU_2009_2}. These kinds of data augmentation techniques
have significantly improved speech recognition accuracy for commercial
products such as Google Home \cite{
c_kim_interspeech_2017_00, B_Li_INTERSPEECH_2017_1, c_kim_interspeech_2018_00}.
However, a still remaining question is what would be the best way to
obtain features as inputs to the neural network.

Using the capabilities of neural networks,
researchers have explored raw-waveform features
\cite{T_Sainath_IEEETran_2017_1}
or complex Fast Fourier Transform (FFT) features
\cite{c_kim_interspeech_2017_00, B_Li_INTERSPEECH_2017_1}.
However, log-mel filterbank coefficients or Mel Filterbank Cepstral Coefficients 
(MFCCs) \cite{pmermelstein1975} still remains the dominant form as 
features of the automatic speech recognition systems 
\cite{h_hadian_interspeech_2018_00, d_amodei_pmlr_2016_00,
c_kim_icassp_2018_01, c_kim_interspeech_2014_00}. This is because the conventional
features such as MFCC or log-mel filterbank coefficients requires
less computation than the neural network-based features 
such as raw-waveform features \cite{T_Sainath_INTERSPEECH_2015_1} while 
showing comparable performance.  In log-mel filterbank coefficients and MFCC, the
log-nonlinearity is employed to represent the relationship
between the perceived sound intensity by human and
the filterbank energy \cite{C_Kim_PhDThesis_2010}. In more recent
features such as Power Normalized Cepstral Coefficients (PNCCs), 
the power-law nonlinearity with the power coefficient of $\frac{1}{15}$
is employed \cite{C_Kim_IEEETran_2016_1, C_Kim_ICASSP_2012_1}. In our previous study 
\cite{C_Kim_ICASSP_2010_1, C_Kim_INTERSPEECH_2009_2} 
this power-law nonlinearity has been shown 
to be more robust against additive noise. Both the log-law nonlinearity
and the power-law nonlinearity with this specific coefficient of $\frac{1}{15}$
 were 
motivated by {\it the rate-intensity relation} of the human auditory system
\cite{C_Kim_PhDThesis_2010, C_Kim_IEEETran_2016_1}.

In this paper, we take a completely different approach. Instead of trying to model
the human auditory system directly, we try to find a nonlinearity function
which maximizes the uniformity of distribution.
We refer this approach to Maximum Uniformity of Distribution (MUD) approach.
This approach is based on the assumption that even though 
neural networks have remarkable capabilities in classifying input features,
training would be easier
 if feature distribution is not too much 
skewed and features are not too much concentrated in an extremely narrow
interval. More specifically, we assume that if the distributions of features 
are difficult to learn, parameter convergence usually becomes more difficult due to the
erratic surfaces of error functions. In this case, we might have hard time 
in fine-tuning learning rates and hyper parameters to obtain converged 
parameters. In this paper, ``easier" training means that the neural network may be
trained well without too much fine-tuning thanks to the well-behaved feature
distribution and the error function surface. 
It has been known that the distribution of amplitudes \cite{C_Kim_INTERSPEECH_2008} and 
filterbank energies  \cite{C_Kim_ASRU_2009_1}
is very sharp and skewed.
Thus, it is usually not possible to use  mel filterbank energy as features
without using any compressive nonlinearity.
We proposed two different types of MUD approaches: power function-based MUD 
and histogram-based MUD. In these approaches, we first obtain the mel filterbank
energy. 
With the power function-based MUD, we apply a power-function based
nonlinearity where the power function coefficient is chosen to maximize the 
likelihood assuming that the nonlinearity output is the uniform distribution.
  With the histogram-based MUD, the empirical Cumulative Density Function (CDF) 
  is obtained from the training database to transform the original distribution
 into a uniform distribution.
In these two approaches, unlike our previous study 
\cite{C_Kim_IEEETran_2016_1, C_Kim_PhDThesis_2010}, we do not use any prior knowledge 
about {\it the rate-intensity relationship} which is the relation between 
the energy of the incoming signal and the perceived intensity 
by a human \cite{C_Kim_PhDThesis_2010}. However, as will be discussed in Sec.  
\ref{sec:end_to_end_speech_recognition}, we may obtain 
surprisingly similar coefficients  to those obtained from human
auditory systems in a {\it data-driven} way.  A major contribution of this 
work is that we developed a new algorithm for designing
the compressive nonlinearity in a {\it data-driven} way, which is much more
flexible than the previous approaches and may be extended to other domains as
well.
Experimental results with an end-to-end speech recognition system
demonstrate that Power-function based MUD shows better result than the 
  conventional Mel Filterbank Cepstral Coefficients (MFCCs) while Histogram-based
  MUD shows comparable results to the MFCC processing. 

The rest of the paper is organized as follows: We develop the theory of
maximizing the uniformity in Sec. \ref{sec:mud_theory}. We describe
the MUD nonlinearity estimation and the entire end-to-end speech
recognition system in Sec. \ref{sec:end_to_end_speech_recognition}.
Experimental results that demonstrates the effectiveness of the MUD processing
is presented in Sec.  \ref{sec:experimental_results}. We conclude in Sec.
\ref{sec:conclusion}.
\section{Maximization of Distribution Uniformity}
\label{sec:mud_theory}
\subsection{Power-function based maximization of distribution uniformity}
\label{sec:power_function_based_mud}
Consider a random variable $\mathbf{X}$ whose range is a closed interval
$I_{\mathbf{X}} = \left[x_{\text{min}}, x_{\text{max}} \right]$. 
$x_{\text{min}}$ and  $x_{\text{max}}$ are the minimum and maximum
values of the random variable $\mathbf{X}$ respectively.

Our objective is to apply a nonlinearity $\sigma_{p}(\cdot)$ 
in the form of \eqref{eq:power_mud} to $X$  so that the transformed
random variable $Y$ closely follows a uniform distribution:
\begin{align}
  \mathbf{Y} = \sigma_{p} \left(\mathbf{X} \right) = (X - x_{\text{min}})^{\alpha}.
    \label{eq:power_mud}
\end{align}
We chose the power function as the nonlinearity, partly because
it has been shown that this function is quite effective as
a compressive nonlinearity in speech feature processing 
\cite{C_Kim_IEEETran_2016_1,C_Kim_ICASSP_2012_1,
C_Kim_ICASSP_2010_1, C_Kim_INTERSPEECH_2009_2}.
We subtract $X$ by $x_{\text{min}}$, since this will simplify
the maximum likelihood estimation of $\alpha$, which will be
explained shortly.
From \eqref{eq:power_mud}, the range of $Y$ is given by
$I_{\mathbf{Y}} = \left[0, \left(x_{\text{max}} - x_{\text{min}} \right)
    ^{\alpha}\right]$. 
Thus, we expect $Y$ to follow the following uniform distribution:
\begin{align}
  \mathbf{Y} \sim 
  \mathcal{U}(0, \left(x_{\text{max}} - x_{\text{min}} \right)^{\alpha} ).
    \label{eq:pdf_y}
\end{align}
The PDF of $Y$ is given by:
\begin{align}
  p_{\mathbf{Y}}(y) =             
    \begin{cases}
      \frac{1}{\left(x_{\text{max}} - x_{\text{min}} \right)^{\alpha}} ,
          & 0 \le y \le \left(x_{\text{max}} - x_{\text{min}} \right)^{\alpha}   \\
      0, \qquad                    & \text{otherwise}.
    \end{cases}
\end{align}
Using the property of the PDFs of the transformed random variables
\cite{A_Papoulis_McGrawHill_2002},
we obtain the PDF of the random variable $\mathbf{X}$ by:
\begin{align}
  p_{\mathbf{X}}(x)
                & = p_{\mathbf{Y}}(y) \frac{dy}{dx}   \nonumber \\
                & = p_{\mathbf{Y}}(y) \left[ \alpha 
                    \left(x - x_{\text{min}}\right)^{\alpha-1} \right]
                    \nonumber \\
                & = 
                \begin{cases}
                  \frac{\alpha \left(x - x_{\text{min}}\right)  ^{\alpha-1}}
                    {\left(x_{\text{max}} - x_{\text{min}}\right)^{\alpha}},  
                      \qquad &  x_{\text{min}} \le x \le x_{\text{max}} \\ 
                  0,         &\text{otherwise}.
                \end{cases}
  \label{eq:pdf_x}
\end{align}
Now, suppose that we have the following $N$ samples from 
the random variable $\mathbf{X}$:
\begin{align}
  X = \left\{x_0, x_1, \cdots, x_{N-1}\right\}.
    \label{eq:x_samples}
\end{align}
Using \eqref{eq:pdf_x}, we obtain the $\alpha$ value which maximizes 
the data likelihood $p(X|\alpha)$.
The log likelihood of the data $X$ assuming the PDF in \eqref{eq:pdf_y}
is given by:
\begin{align}
  \mathcal{L}\left(\alpha ; X\right) 
      & = \sum_{i=0}^{N-1} \ln p_{\mathbf{X}} \left( x_i \right)  \nonumber \\
      & = \sum_{i=0}^{N-1} 
              \ln \left[ \frac{ \alpha (x_i - x_{\text{min}} )^{\alpha-1}}  
              {\left(x_{\text{max}} - x_{\text{min}} \right)^{\alpha}} \right]  \nonumber \\
      & = N \ln(\alpha) + \left(\alpha-1 \right) \sum_{i=0}^{N-1}
        \ln \left(x_i - x_{\text{min}} \right)  \nonumber \\
      & \qquad - N \alpha \ln \left(x_{\text{max}} - x_{\text{min}} \right).
        \label{eq:x_likelihood}
\end{align}
In \eqref{eq:x_likelihood}, the term $\ln \left(x_i - x_{\text{min}}
\right)$ is not defined when $x_i = x_{\text{min}}$. Thus, we apply flooring as
shown below:
\begin{align}
  \mathcal{L}\left(\alpha ; X\right) & =  
    N \ln(\alpha) \nonumber \\
     & \quad  + \left(\alpha-1 \right) \sum_{i=0}^{N-1}
        \ln \left(\max \left\{x_i - x_{\text{min}}, \delta \right\} \right)
          \nonumber \\
      & \quad - N \alpha \ln \left(x_{\text{max}} - x_{\text{min}} \right),
        \label{eq:x_likelihood}
\end{align}
where $\delta$ is a flooring coefficient. We use $\delta = 10^{-100}$ in our
experiments.
By differentiating $\mathcal{L}\left(\alpha | X\right)$ with respect to 
$\alpha$, we obtain $\hat{\alpha}$, which maximizes the likelihood as below:
\begin{align}
  \hat{\alpha} = \frac{1}{
    \ln \left(\max \left\{x_i - x_{\text{min}}, \delta \right\} \right)
    - \frac{1}{N} \sum_{i=0}^{N-1} \ln \left(x_i - x_{\text{min}}\right)}.
  \label{eq:alpha_estimation}
\end{align}
\subsection{Histogram-based maximization of distribution uniformity}
\label{sec:histogram_based_mud}
Instead of using the power-function based parametric approach to
maximize the uniformity of distribution, we may also consider the
non-parametric approach. In this approach, we estimate the 
Cumulative Distribution Function (CDF) from the samples in 
\eqref{eq:x_samples}. This CDF estimation is achieved by 
sorting the samples $x_i$ in \eqref{eq:x_samples} and performing
interpolation. The relation between the original random variable
$\mathbf{X}$ and the transformed random variable $\mathbf{Y}$
is given by the following equation:
\begin{align}
  \mathbf{Y} = \sigma_{np}(\mathbf{X}) 
      = F_u^{-1} \left(\hat{F}_x(\mathbf{X})\right)
  \label{eq:histogram_mud}
\end{align}
where $F_u(\cdot)$ is the CDF of the uniform distribution.
$\hat{F}_x(\mathbf{X})$ in \eqref{eq:histogram_mud} is the 
estimated CDF of $\mathbf{X}$ that is mentioned above.
In the special case of the uniform distribution of 
$\mathcal{U}(0, 1)$, the inverse of this CDF 
is given by  $F_u^{-1}(x) = x, \;\; 0 \le x \le 1$. Under this
assumption, the above equation \eqref{eq:histogram_mud} may be 
simplified to:
\begin{align}
  \mathbf{Y} = \sigma_{np}(\mathbf{X}) 
      = \hat{F}_x(\mathbf{X}).
  \label{eq:histogram_mud_simple}
\end{align}

\section{End-to-End speech recognition with the maximization of feature 
distribution uniformity}
\label{sec:end_to_end_speech_recognition}
\begin{figure}[tbp]
  \begin{center}
    \resizebox{80mm}{!}{\input{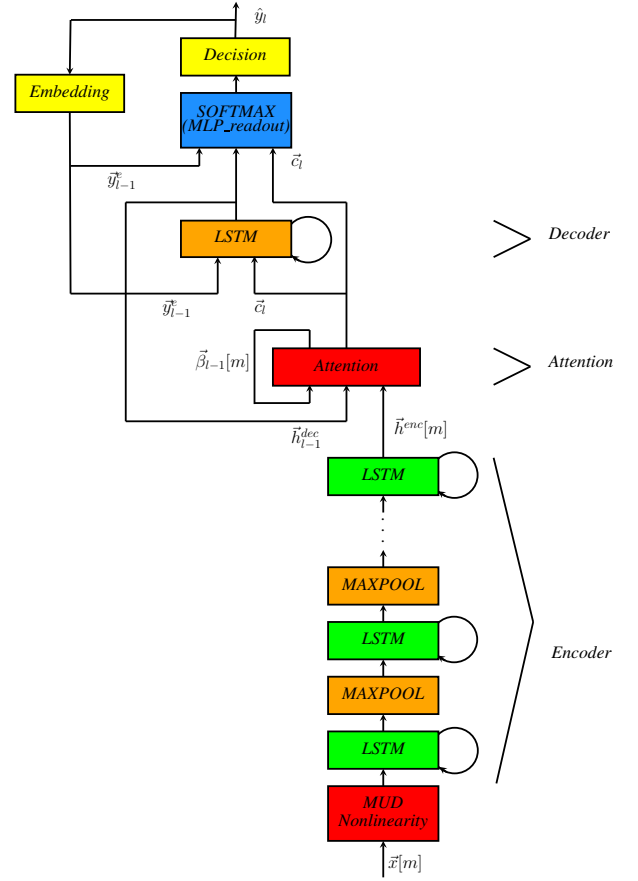}}
      \caption {  The structure of the entire end-to-end 
      speech recognition system with MUD processing. The LSTMs
      in the encoder layers may be either bidirectional-LSTMs or
      unidirectional-LSTMs. The attention may be either the full
      attention or the MOnotonic CHunkwise Attention (MoCha) 
      \cite{c_chiu_iclr_2018_00, k_kim_asru_2019_00}. 
     }   
     \label{fig:entire_diagram}
  \end{center}
\end{figure}
In this section, we explain how to use  the theories we developed in
Sec. \ref{sec:power_function_based_mud} and \ref{sec:histogram_based_mud}
to train an end-to-end speech recognition system. The entire block diagram
of the system is shown in Fig. \ref{fig:entire_diagram}. 
We used two different attention structures: Bidirectional LSTMs with 
Full Attention (BFA)  \cite {d_bahdanau_iclr_2015_00} and 
MOnotonic CHunkwise Attention (MoCha) 
\cite{c_chiu_iclr_2018_00}. Our MoCha implementation is described
in very detail in our another paper \cite{k_kim_asru_2019_00}.

We apply either the 
power function-based MUD nonlinearity in \eqref{eq:power_mud} or the 
histogram-based MUD nonlinearity to each mel filterbank energy as the first
step as depicted in Fig. \ref{fig:entire_diagram}.
The mel filterbank energy is defined by the following equation:
\begin{align}
  p[m, l] & = 
    \sum_{k=0}^{K/2} \left|X[m, e^{j \omega_k}]\right|^2  M_l[\omega_k] 
  \label{eq:mel_filterbank_energy}
\end{align}
where $M_l[\omega_k]$ is the triangular mel response for the $l$-th
filterbank channel, $m$ is the frame index, and $K$ is the Fast Fourier
Transform (FFT) size. $\omega_k$ is the discrete-time frequency defined
by $\omega_k = \frac{2 \pi k}{K} \;\; 0 \le k \le K - 1$. 
The input feature vector $\vec{x}[m]$ in 
Fig. \ref{fig:entire_diagram} is therefore given by:
\begin{align}
  \vec{x}[m] = \left[p[m, 0], \; p[m, 1], \; \cdots, \; p[m, C-1] \right].
\end{align}
where $C$ is the number of mel filter bank channels. In our experiments, we
used the value of $C = 40$. For the power function-based MUD, we use 
\eqref{eq:alpha_estimation} for each mel filterbank channels from the 
randomly selected 1,000 utterances from the training set. In order not to
be affected by the silence portion, we removed non-speech portion using
a simple energy-based Voice Activity Detector (VAD). 

\begin{figure}[tbp]
  \begin{center}
    \includegraphics[width=70mm]{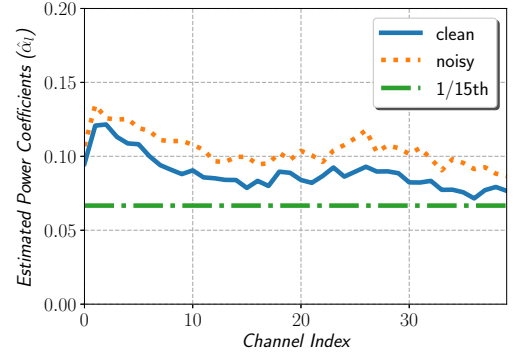} 
    \caption {
      \label{fig:estimated_power_coeff}
      The estimated power coefficients for each mel filterbank channels using
      \eqref{eq:alpha_estimation}.
    }
  \end{center}
  \vspace{-7mm}
\end{figure}

Fig.  \ref{fig:estimated_power_coeff} shows the estimated $\hat{\alpha}$ 
using \eqref{eq:alpha_estimation} for each mel filterbank channel.
From this Fig. \ref{fig:estimated_power_coeff}, we observe that 
$\hat{\alpha}$ values are surprisingly close to the power 
coefficient of $\frac{1}{15}$ which we obtained by modeling the rate-intensity 
curve using a human auditory system \cite{C_Kim_PhDThesis_2010,
C_Kim_IEEETran_2016_1}. 
For the histogram-based
MUD, we also used the same randomly selected 1,000 utterances from the 
training set, applied a VAD, and constructed the empirical CDF to
obtain the nonlinearity function in \eqref{eq:histogram_mud}.

\begin{figure}[tbp]
  \begin{center}
    \subfloat[] {
      \includegraphics[width=70mm]{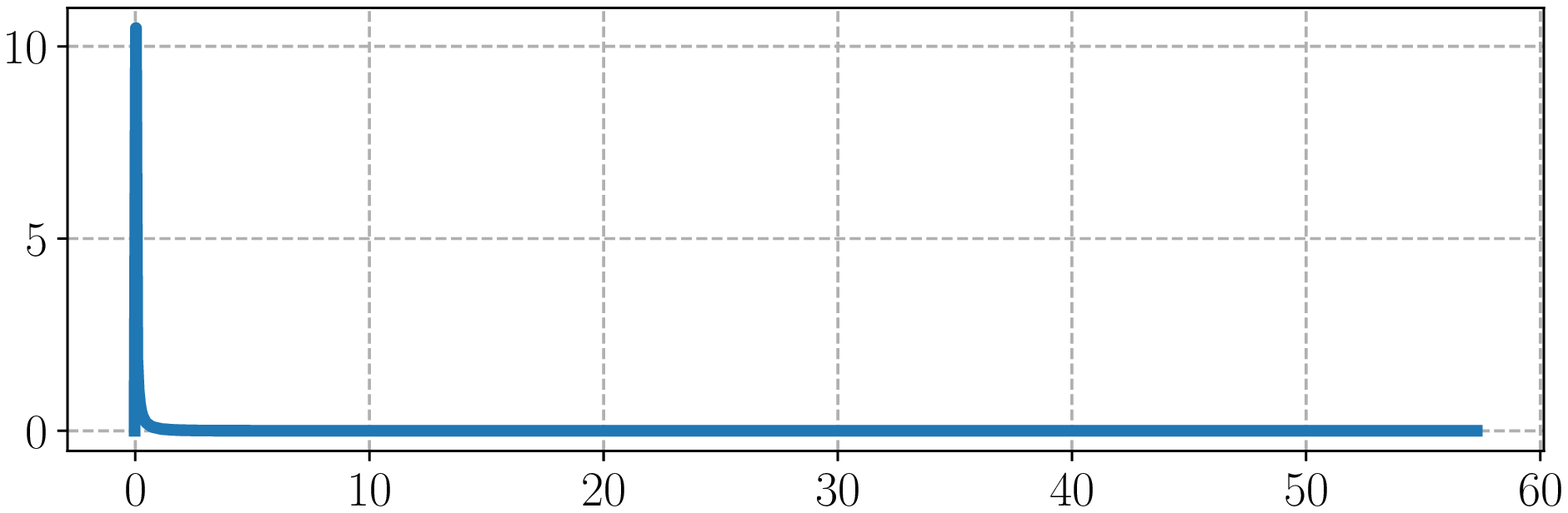}
      \label{fig:mel_filterbank_energy_histogram}
  \vspace{-2mm}
    }   
    \\  
    \subfloat[] {
      \includegraphics[width=70mm]{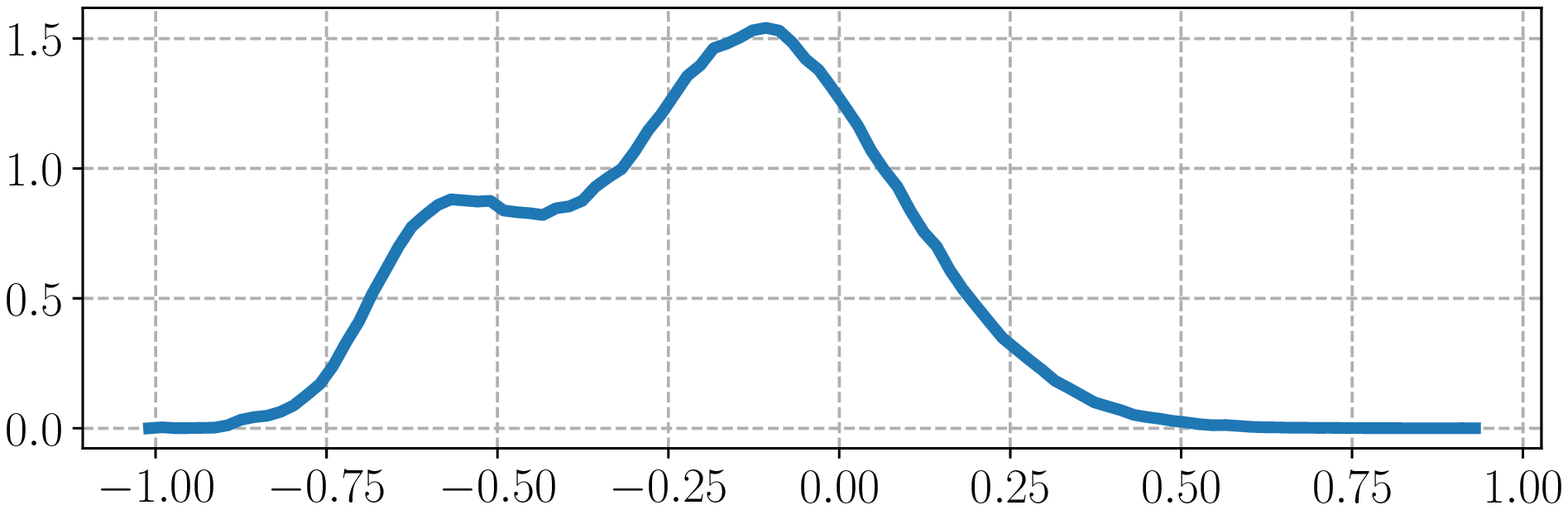}
      \label{fig:power_function_based_mud_output_histogram}
    }   
  \vspace{-2mm}
		\\
    \subfloat[] {
      \includegraphics[width=70mm]{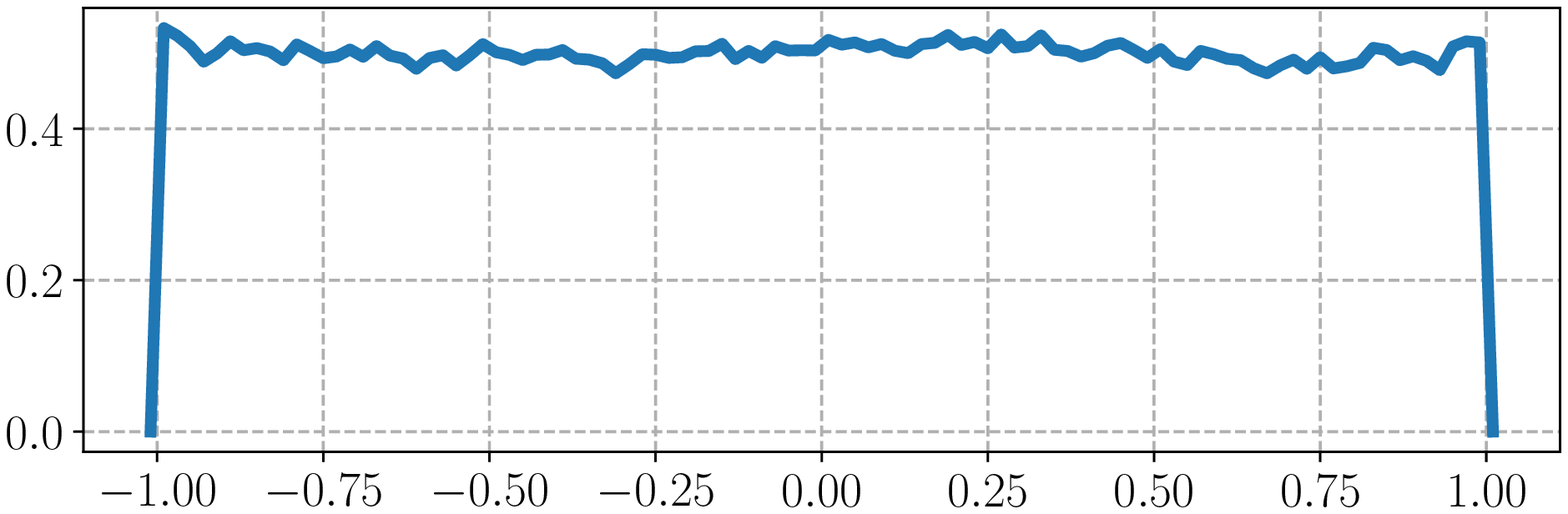}
      \label{fig:histogram_based_mud_output_histogram}
    }   
  \vspace{-2mm}
    \caption {
      The Probability Density Functions for the third filterbank channel
       of (a): the mel filterbank energy $p[m, l], l = 3$ in 
       \eqref{eq:mel_filterbank_energy}, (b): the power-function based MUD
      output of this mel filterbank energy in \eqref{eq:power_mud}, (c)
      the histogram based-MUD in \eqref{eq:histogram_mud}. 
      \label{fig:pdf_mud}
    }   
   \vspace{-7mm}
  \end{center}
\end{figure}

\begin{figure}[tbp]
  \begin{center}
    \includegraphics[width=70mm]{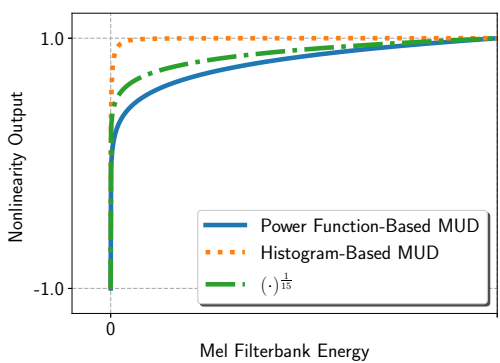}
      \caption { 
      \label{fig:nonlinearity_comparison}
      Comparison of different nonlinearities:
      Power-law nonlinearity of the form of $(\cdot)^{\frac{1}{15}}$
      used in PNCC \cite{C_Kim_IEEETran_2016_1, C_Kim_ICASSP_2012_1},
      power function-based MUD in \eqref{eq:power_mud}, 
      and histogram-based MUD \eqref{eq:histogram_mud} for the
      third mel filterbank channel. 
      }
  \vspace{-7mm}
  \end{center}
\end{figure}

Fig. \ref{fig:pdf_mud} shows the Probability Density Functions (PDFs)
of the  mel filterbank energy in (a), those of the nonlinearity output using 
the power-based MUD in (b), and those of the nonlinearity output
using the histogram-based MUD in (c). In plotting these PDFs in Fig.
\ref{fig:pdf_mud}, we used another 1,000 utterances which are not 
included in estimating the MUD nonlinearities. These plots are for the
third filterbank channel $l=3$ in \eqref{eq:mel_filterbank_energy}.
As shown in Fig. \ref{fig:power_function_based_mud_output_histogram}, 
if we use the power function-based MUD, the PDF becomes much smoother
compared to the original PDF in Fig. \ref{fig:mel_filterbank_energy_histogram}.
However, this PDF is not as uniform as the one in Fig.  
\ref{fig:histogram_based_mud_output_histogram}. In Fig.
\ref{fig:nonlinearity_comparison}, we compared the Power-law nonlinearity 
of the form of $(\cdot)^{\frac{1}{15}}$ used in PNCC
\cite{C_Kim_IEEETran_2016_1, C_Kim_ICASSP_2012_1},
power function-based MUD in \eqref{eq:power_mud}, 
and histogram-based MUD \eqref{eq:histogram_mud} for the
third mel filterbank channel $l=3$ in \eqref{eq:mel_filterbank_energy}. 
Note that in case of power function-based MUD and histogram-based MUD, the nonlinearity
functions are different for different filterbank channels.
\begin{table*}[!htbp]
  \renewcommand{\arraystretch}{1.3}
  \centering
      \caption{\label{tbl:result} 
      Word Error Rates (WERs) obtained with MFCC, Power Mel filterbank
      coefficients, \\ power function-based MUD processing, and
      histogram-based MUD Processing on the LibriSpeech corpus 
      \cite{v_panayotov_icassp_2015_00}. \\ For each WER number, the same 
      experiment was conducted twice and the results were averaged. \\ All these 
      results were obtained \textbf{without} using a Language Model (LM).
      }
      \begin{tabular}{| c | c  || c | c | c | c |}
        \hline
        \multicolumn{2}{| l ||}{Neural Network Structure}
                               & \specialcell{MFCC}
                               & \specialcell{$ ( \cdot ) ^ {\frac{1}{15}}$}
                               & \specialcell{Power Function- \\ Based MUD}
                               & \specialcell{Histogram- \\ Based MUD}   \\
        \hline \hline
        \multirow{3}{*}{\specialcell{1024 cell \\ ULSTM \\ MoCha}}    
                & test-clean  &   7.09  \% & \textcolor{blue}{\textbf{ 7.04 }} \%   &   7.10 \%  &  7.13 \%  \\
                & test-other  &  20.60  \% & 19.76 \%
                &  \textcolor{blue}{\textbf{19.64}} \%  & 20.03 \%  \\
                & average     &  13.85  \% & 13.40 \%
                &  \textcolor{blue}{\textbf{13.37}} \%  & 13.58 \%  \\
        \hline \hline
        \multirow{3}{*}{\specialcell{1536 cell \\ BLSTM \\ Full-Attention}}
                & test-clean  &   4.06  \% & \textcolor{blue}{\textbf{3.94}} \%   &   4.02 \%  &  4.11 \%  \\
                & test-other  &  13.97  \% & 13.56 \%
                & \textcolor{blue}{\textbf{13.34}} \%  & 14.10 \%  \\
                & average     &   9.02  \% &  8.75 \%
                & \textcolor{blue}{\textbf{8.68}} \%  &  9.11 \%  \\
        \hline
     \end{tabular}
     \vspace{-4mm}
\end{table*}
We used the RETURNN speech recognition system \cite{p_doetsch_icassp_2017_00,
a_zeyer_interspeech_2018_00, c_luscher_interspeech_2019_00}.
We have tried various modifications to the training stratgegy
(\emph{e.g} \cite{d_gowda_interspeech_2019_00, a_garg_asru_2019_00}).
$\vec{x}[n]$ and 
$\vec{y}_l$ are the input mel filterbank energy vector and the output label
, respectively. $m$ is the input frame index and $l$ is the decoder output
step index. $\vec{c}_l$ is the attention context vector calculated as
a weighted sum of the encoder hidden state vectors $\vec{h}_{enc}[m]$. The 
weights used in this procedure is called the attention weights. They are
calculated by applying softmax to the attention energies 
\cite{d_bahdanau_iclr_2015_00, a_zeyer_interspeech_2018_00}.
$\vec{h}^{enc}[m]$ and $\vec{h}^{dec}_l$ are the encoder and the decoder
hidden state vectors, respectively. $\vec{\beta}_l[m]$ is the attention weight
feedback \cite{a_zeyer_interspeech_2018_00}.
In \cite{a_zeyer_interspeech_2018_00}, the peak value of the speech 
waveform is normalized to be one. However, since finding the peak sample value 
is not possible for on-line feature extraction, we did not perform this
normalization. We modified the input pipeline
so that the on-line feature generation can be performed. We disabled the
clipping of feature range between -3 and 3, which is the default setting in 
their LibriSpeech experiment in \cite{a_zeyer_interspeech_2018_00}. 
We conducted experiments using both the
uni-directional and bi-directional Long Short-Term Memories  (LSTMs)
\cite{S_Hochreiter_neural_computation_1997_00}. 
For on-line processing, we used the MOnotonic CHunkwise Attention 
(MoCha) \cite{c_chiu_iclr_2018_00}. 
In online speech recognition experiments using MoCha, we used the chunk
size of 2.
 For better 
stability in the LSTM training, we used the gradient clipping by 
global norm \cite{r_pascanu_icml_2013}, which is implemented as 
{\tt tf.clip\_by\_global\_norm } API in Tensorflow  \cite{m_abadi_usenix_2016}.
We used six layers of encoders and one layer of decoder followed by a softmax 
layer. The training infrastructure we used is described in more detail in 
our another paper \cite{c_kim_asru_2019_01}.
\section{Experimental Results}
\label{sec:experimental_results}

\label{sec:experimental_results}
For speech recognition experiments, we used the Librispeech database
\cite{v_panayotov_icassp_2015_00}
for training and evaluation. For training, we used the entire 960 hours training
set consisting of 281,241 utterances. For evaluation, we used the 
official 5.4 hours {\tt test-clean} and 5.1 hours {\tt test-other} databases.
We conducted experiments
using the 40-th order MFCC feature implemented in
\cite{b_mcfee_proc_scipy_2015_00}, power-law nonlinearity of
$(\cdot)^{\frac{1}{15}}$ applied to the mel filterbank energy, 
power function based MUD processing, and histogram-based MUD processing.
We conducted experiments using both the online ULSTM/MoCha 
\cite{c_chiu_iclr_2018_00} structure and the BLSTM with the full-attention 
structure. We have conducted Bidirectional Long Short-Term Memory (BLSTM) 
experiments with the cell size of 1536.
For the on-line MoCha experiment, we used the Uni-directional 
Long Short-term Memory (ULSTM) with the cell size of 1024. 
Our MoCha implementation is described in very detail in \cite{k_kim_asru_2019_00}.
In all of our experiments in this section, we did not use any external Language 
Models (LMs). We observed that external LMs can significantly enhance the 
speech recognition accuracy of our end-to-end speech recognition system, which 
is shown in our other papers
\cite{c_kim_interspeech_2019_00, c_kim_asru_2019_01}. However, in this paper,
just to focus on the effects of nonlinearity, we did not employ external LMs.

These results are summarized in Table \ref{tbl:result}. For
each WER number in this table, the same experiment was conducted twice and 
these results were averaged to reduce the effect of random fluctuation
in each trial.
The best 
performance was achieved when we used the power function-based MUD with 
the 1536-cell BLSTM layers in the encoder and the full attention.
For the {\tt test-clean} and test {\tt test-other} test sets
\cite{v_panayotov_icassp_2015_00} , we 
obtained 4.02 \% Word Error Rate (WER) and 13.34 \% WER, respectively.
On average, the WER was 8.68 \%, which is relatively 3.77 \% improvement   
over the baseline MFCC with 9.02 \% WER.
From Table. \ref{tbl:result},
we note that usually there is no improvement over the baseline MFCC on
the {\tt test-clean} set. However, improvement on the {\tt test-other}
was usually more substantial. For the 1536-cell BLSTM full-attention case,
the relative improvement over the baseline MFCC on the {\tt test-other} 
is 4.51 \%.
 The performance difference between
the power-law nonlinearity of $(\cdot)^{\frac{1}{15}}$ and the power
function-based MUD is usually very small. This was expected since
the estimated parameters using \eqref{eq:alpha_estimation} are not
very different from $\frac{1}{15}$ as shown in Fig.
\ref{fig:estimated_power_coeff}.  However, for the {\tt test-other}
database, which is more a difficult set, the improvement over 
the power-law nonlinearity of $(\cdot)^{\frac{1}{15}}$ is {\bf statistically
significant}.
Histogram-based MUD shows somewhat worse 
performance compared to power function-based MUD. However, this histogram-
based MUD still shows comparable results to the conventional MFCC processing.
For the histogram-based MUD, we also tried to transform the PDF into a Gaussian
distribution. However, that system showed slightly worse results than the 
Histogram-based MUD in Table \ref{tbl:result}.
We hypothesize that the reason why histogram-based MUD does slightly worse
than the power-function based MUD is that it somewhat obscured the
energy boundary between speech vs non-speech. If we can employ a very sharp VAD
to select only the speech portion very accurately, we think the performance
of the histogram-based MUD will be comparable to that of the power-funcation
based MUD.

\section{Conclusions}
\label{sec:conclusion}
In this paper, we described the Maximum Uniformity of Distribution 
(MUD) algorithm. This approach is based on the assumption that
neural-network training would be easier and the converged parameters would show better
performance when feature distribution is not too much 
skewed or too much concentrated in an extremely narrow interval. We proposed
two different types of MUD approaches: power function-based MUD and histogram
-based MUD. In these approaches, we first obtain the Mel filterbank
coefficients. The estimated parameters using the power function-based MUD 
using \eqref{eq:alpha_estimation} are surprisingly close to the power 
coefficient of $\frac{1}{15}$ which we obtained by modeling the rate-intensity 
curve using a human auditory system \cite{C_Kim_PhDThesis_2010,
C_Kim_IEEETran_2016_1}. The histogram-based MUD shows comparable performance
to the conventional MFCC processing, but it was worse than the performacne
of the power function-based MUD. In the end-to-end speech recognition
experiments on the LibriSpeech databases \cite{v_panayotov_icassp_2015_00},
we obtained 4.02 \%  WER and 13.34 \% WER on the  {\tt test-clean} and 
test {\tt test-other} test sets respectively using the power function-based
MUD processing. 

The major novelty of this paper is that we proposed a new way of deriving 
a suitable nonlinearty from the training data themselves in a 
{\it data-driven} way. 
In the case of the previous power-law nonlinearty with the power coefficients of 
$\frac{1}{10}$ \cite{C_Kim_INTERSPEECH_2009_2} or $\frac{1}{15}$ \cite{C_Kim_ICASSP_2012_1}, 
they were obtained either by curve-fitting from {\it the rate-intensity curve } of the 
human auditory system \cite{C_Kim_IEEETran_2016_1}
or by performing speech recognition experiments with various power coefficients 
to find out the optimal value \cite{C_Kim_PhDThesis_2010}. 
Since these steps require significant amount time, 
we could not fine-tune the power coefficient for 
each filter bank channel in our previous work \cite{C_Kim_IEEETran_2016_1}. 
In this new MUD approach, we obtain suitable coefficients for 
each filterbank channel ``without" actually running speech recognition experiments. 
This is a significant advantage compared to the previous hand-crafted fine-tuning. 
In addition, we believe that this approach is not only limited to speech
recognition, but it can be applied to other domains in the future. 
Since the  previous power coefficient of $\frac{1}{15}$ was already hand-optimized 
by performing experiments with different coefficients, it is quite natural 
that the additional improvement of MUD over the previous power-law nonlinearity
is relatively small, which is also shown in Section 
\ref{sec:experimental_results}.
Nevertheless, the major contribution of this work is that we proposed 
a new way of designing the compressive nonlinearity in a {\it data-driven } way,
and this approach is much more flexible and may be extended to other domains 
as well.
%
%
%
%
\ninept
\bibliographystyle{IEEEtran}
\bibliography{main}

\end{document}